\documentclass[a4paper,fleqn]{cas-dc}

\usepackage[authoryear]{natbib}
\usepackage{booktabs}
\usepackage{makecell}
\usepackage{float}
\usepackage[flushleft]{threeparttable}

\usepackage{listings}
\usepackage{xcolor}
\usepackage{array}
\usepackage{multirow}
\def\tsc#1{\csdef{#1}{\textsc{\lowercase{#1}}\xspace}}
\tsc{WGM}
\tsc{QE}
\tsc{EP}
\tsc{PMS}
\tsc{BEC}
\tsc{DE}

\begin{document}
\let\WriteBookmarks\relax
\def\floatpagepagefraction{1}
\def\textpagefraction{.001}

\shorttitle{Marine Digital Twin Platform}

\shortauthors{Yu Ye et~al.}

\title [mode = title]{Advancing Towards a Marine Digital Twin Platform: Modeling the Mar Menor Coastal Lagoon Ecosystem in the South Western Mediterranean}                      
\tnotemark[1,2]



%




\affiliation[1]{organization={Department of Information and Communication Engineering},
    addressline={University of Murcia}, 
    city={Murcia},
    postcode={30100}, 
    country={Spain}}

\affiliation[2]{organization={Department of Ecology and Hydrology, Faculty of Biology, and Regional Campus of International Excellence ``Mare Nostrum”},
    addressline={University of Murcia}, 
    city={Murcia},
    postcode={30100}, 
    country={Spain}}

\author[1]{Yu Ye}[
                orcid=0000-0002-6744-3338]

\author[1]{Aurora González-Vidal}[
                orcid=0000-0002-4398-0243]
\cormark[1]
\ead{aurora.gonzalez2@um.es}

\author[1]{Alejandro Cisterna-García}[
                orcid=0000-0002-9927-7178]


\author[2]{Angel Pérez-Ruzafa}[
                orcid=0000-0003-4769-8912]

\author[1]{Miguel A. Zamora Izquierdo}[
                orcid=0000-0002-0130-2364]                

\author[1]{Antonio F. Skarmeta}[
                orcid=0000-0002-5525-1259]                

\cortext[cor1]{Corresponding author}



\begin{abstract}
Coastal marine ecosystems face mounting pressures from anthropogenic activities and climate change, necessitating advanced monitoring and modeling approaches for effective management. This paper pioneers the development of a Marine Digital Twin Platform aimed at modeling the Mar Menor Coastal Lagoon Ecosystem in the Region of Murcia. The platform leverages Artificial Intelligence to emulate complex hydrological and ecological models, facilitating the simulation of \textit{what-if} scenarios to predict ecosystem responses to various stressors. We integrate diverse datasets from public sources to construct a comprehensive digital representation of the lagoon's dynamics. The platform's modular design enables real-time stakeholder engagement and informed decision-making in marine management. Our work contributes to the ongoing discourse on advancing marine science through innovative digital twin technologies.
\end{abstract}


\begin{highlights}
\item Development of a Marine Digital Twin Platform to model the Coastal Ecosystem of the Mar Menor Lagoon in the Region of Murcia.
\item Integration of Artificial Intelligence to emulate complex hydrological models and facilitate simulations of hypothetical scenarios.
\item Modular design of the platform that allows real-time stakeholder engagement and informed decision-making in marine management.
\end{highlights}


\maketitle

\section{Introduction} 

The intricate dynamics of coastal marine ecosystems present significant challenges for both monitoring and understanding their processes. Oceans are vital for sustaining life on Earth and they contribute substantially to global food sources, oxygen production, and carbon dioxide absorption \citep{riebesell2009sensitivities}. 
Marine environments suffer from numerous sources of stress, mostly from human activities in coastal areas, urban, agricultural, and industrial discharges, habitat destruction, introduction of invasive species, and oil spills, which interact synergistically with the consequences of climate change. In addition to classic pollutants, such as heavy metals or pesticides, with a long tradition in human activities such as mining, industry, or agriculture, new emerging pollutants are continually appearing, derived from drugs or cosmetics, whose effects on health are not always well known. A good example is that of microplastics \citep{ritchie2018plastic}.
These pressures result in the alteration of ecological processes, eutrophication and dystrophic crises, anoxia, pH alterations, changes in community structure, the collapse of populations of some species and massive proliferation of others, and accumulation of toxic compounds through food webs. This impacts not only human health but also the habitability of coastal areas, biodiversity, the sustainability of resource exploitation, leading to important socio-economic consequences.


Given this scenario, the undersampling of marine environments, particularly in continental shelves and coastal areas, frustrates our comprehensive understanding of marine sciences \citep{davidson2019synergies}, since such complex systems require continuous monitoring of various indicators to detect or alert us to changes. Current observational deployments are often restricted to the ocean surface and a few measurable variables and there are limited tools to process the data and extract useful knowledge. This underscores the need for advanced modeling techniques to bridge gaps in our comprehension and to allow intelligent action-taking. But, more importantly, the mere detection of problems may not be sufficient since, on the one hand, the homeorhetic mechanisms of biological systems may mask such indicators until it is too late and, on the other hand, the speed of ecosystem deterioration is often greater than the human capacity to take corrective and management measures.

Marine ecosystem models play a pivotal role in conceptualizing relationships among marine organisms and in the understanding of their impact on the environment \citep{ford2018marine}. These models facilitate the calculation of intricate phenomena, offering valuable insights into ecosystem processes. However, challenges persist in the accuracy of unobserved model outputs due to inherent uncertainties and biases, limiting the reliability of predictions. Recognizing these limitations, the concept of Digital Twins of the Ocean (DTOs) emerges as a transformative approach to address the shortcomings of traditional models \citep{skakala2023future, tzachor2023digital}.

DTOs are digital representations \citep{flynn2022plankton} tailored for real-time stakeholder needs and present an adaptive and learning framework that can evolve with new knowledge and observations \citep{ford2022solution}. This innovation holds immense promise for stakeholders, including non-scientific backgrounds, allowing for informed decision-making in marine management. 

In this context, our paper pioneers the creation of a modular DTO that seamlessly uses Artificial Intelligence (AI) and software engineering (especially web applications) to emulate complex hydrological and ecological models and propose \textit{what-if} scenarios. This effort aims to merge technological advancements with marine sciences for a transversal approach to environmental management. Coastal lagoons, particularly productive and sensitive to climate change, serve as ideal ecosystems for studying these interactions.

The primary objective of our DTO is integrating advanced Artificial Intelligence algorithms and gathering the necessary data to emulate complex hydrological models and simulate hypothetical scenarios, enabling real-time stakeholder engagement and informed decision-making.

Precisely, the contributions towards this general goal of our work are:
\begin{itemize}
    \item Identification, fusion, and incorporation of open data sources related to the Mar Menor basin, including biochemical parameters and satellite images for improving their availability and usability.
    \item Development of technologies that compose the DTO in all stages: data collection and fusion, AI model implementation and output generation (back-end) as well as a friendly user interface (front-end).
    \item Design, implement and continuously integrate data into a FIWARE-based application to enable data interoperability.
    \item Outlining potential scenarios addressing issues like jellyfish blooms, hypoxic crisis and water quality estimation highlighting the need of a common open data space.
\end{itemize}

The paper is organized as follows. Section 2 reviews related projects and existing work on marine-related Digital Twins. Section 3 presents the methodology for creating the Digital Twin, covering the architecture, data collection, fusion, and analysis. Section 4 provides a detailed description of the implemented AI models. Section 5 explores several perspectives for the developed Digital Twin, highlighting potential integrations with additional data. Finally, Section 6 concludes the paper and outlines future directions.

\section{Related work}

\subsection{Related studies and projects}
As an emerging and still evolving technology, DTs have experienced an extension from the use in the manufacturing and asset-heavy industries and have made it into complex fields such as urban planning, precision healthcare, and marine science \citep{bronner2023digital}. 

Most marine DT-related work can be divided into two types: tangible, which refers to real objects like power generators, ships, vessels, or buildings; and intangible, which involves natural processes such as natural phenomena at a particular location, global climate change, or even the entire Earth \citep{tzachor2023digital, minerva2020digital}. It is worth noting that the methodological aspects of a DT approach to the study of the Earth system or its sub-components have not yet been established \citep{pillai2022digital}. However, knowledge can be gained from specific application cases through a variety of methods, which can be either physical/numerical models or ML based.

As tangible DT examples, \citep{su2023towards} implemented and tested a full-scale DT solution for an aquaculture net cage system. They developed a wireless sensor network (WSN) and included a set of the most commonly used sensors and technologies (e.g. metocean buoy, online metocean forecast, 4G modules, etc.) for real-time monitoring. They also used FhSim as a simulation framework to combine the different models such as fish behavior or marine environment. To address the problem of estimating the speed loss caused by the effect of fouling on the ship's hull and propeller, \citep{coraddu2019data}  proposed a two-step Data-Driven Model (DDM). The DT is built in the first step using the Deep Extreme Learning Machines (DELM) and leveraging the information collected from the on-board sensors. Then, in the second step, the same model is applied to estimate the speed loss of the ship and its drift. In the work of \citep{stadtmann2023demonstration}, a DT of a floating offshore wind turbine is created. The implemented DT was categorized as standalone, descriptive, and predictive level (on a scale of 0 to 5: 0-standalone, 1-descriptive, 2-diagnostic, 3-predictive, 4-prescriptive, and 5-autonomous). The CAD modeling and texturing, Unity and Virtual Reality (VR) technologies were used to visualize the virtual turbine through an Oculus VR headset, together with the real sensor data recorded on the turbine. Finally, for predictive capability, Dense Neural Networks (DNN) and Long-Short Term Memory Neural Networks (LSTM) were used as ML approaches to predict turbine-specific parameters.

On the other hand, as intangible examples, \citep{duque2022building} implemented a web application as a proof of concept of applicability of data integration in the DTs. It integrates Copernicus and WorldPop data to provide tools for analyzing and describing coastal interactions between ocean, land, and demographic variables on the Italian coasts, providing both visualization and analysis capabilities. Another proof of concept experiment was carried out by \citep{jiang2021digital} with the application called \textit{CoastalTwin}, which uses a physics-informed ML technique called Fourier Neural Operator (FNO) to predict the sea surface height and it showed a great potential as a promising alternative compared to the commonly used Nucleus for European Modeling of the Ocean (NEMO) method, achieving 45 times of acceleration. \citep{ahmadi2024supervised} developed a multi-regional terrain and elevation map by combining a digital twin model and the DL technique (U-net network) on the Florida coast. They scaled 5000 map segments worldwide, each with its main geographical features, and classified the terrain into seven different classes. The implemented DT can serve as both a physical and simulation model, and can also be used in environmental monitoring and urban planning.

Regarding marine DT-related projects, on a larger scale, the Digital Twins of the Ocean (DITTO) project (\url{https://ditto-oceandecade.org/digital-twins/}) aims to create accurate representations of the worldwide marine assets and systems. It establishes a comprehensive digital framework that integrates marine data, satellite observations, modeling, AI, and high-performance computing, allowing users and partners to explore \textit{what-if} questions based on shared data, models and knowledge. At the European level, an infrastructure built on existing resources and initiatives, called the European Digital Twin of the Ocean (EU DTO) (\url{https://www.mercator-ocean.eu/en/digital-twin-ocean/}), aims to model the ocean's multiple components, provide knowledge and understanding of its past and present, and generate reliable predictions of its future behavior. It includes other projects such as: Iliad (\url{https://ocean-twin.eu/}), which pilots local digital twins with the goal of eventually creating a unified virtual replica; the EDITO-Infra project (\url{https://edito-infra.eu/}), which integrates the Copernicus Marine Service, Copernicus Data and Information Access Services, and the European Marine Observation and Data Network; the EDITO-Model Lab project (\url{https://edito-modellab.eu/}), which develops the underlying models for the DT. 
 At local scale, the SmartLagoon project (\url{https://www.smartlagoon.eu/}), related to Mar Menor and funded by the European Union's Horizon 2020 program, aims to develop tools for monitoring and analyzing the socio-environmental interrelationships of coastal lagoons and their ecosystems. Its goal is to raise awareness of environmental impacts at local and citywide levels. Lastly, the objectives of the national project in which this work is framed, ThinkInAzul consist of points that are aligned with the generation of a digital twin: \begin{itemize}
    \item Creation of new digital tools for the observation of the marine environment
    \item Creation of an open platform to monitor and access marine data
    \item Generation of a system, based on observations, data analysis, and numerical models, that allows mechanisms for early warning against risks and threats, resource management, land planning, and the management of the marine environment in its broadest sense
\end{itemize}
ThinkInAzul is part of a joint national research and innovation strategy to protect marine ecosystems against climate change and pollution and to address the challenges related to aquaculture, fishing, and tourism. The Region of Murcia acts as the national coordinator for the Supplementary R\&D\&I Plan in Marine Sciences.

\subsection{Research gaps}
Most DT-related studies often focus on a single use case, with models and data formats that are often not interoperable. This limits the ability to integrate different components into a cohesive system. Our DT is able to support multiple models, each with its own use case, and the implemented data management allows the stored data to be interoperable between different scenarios or even with other systems. 

Compared to the aforementioned projects, despite their extensive data coverage, most of them are interpolated and lack local accuracy. Also, the analysis of the Mar Menor Coastal Lagoon Ecosystem requires more data on the land and the coast, while these data are mainly focused on ocean variables. In addition, as mentioned above, the technology and scenario of DT is still under development, the use cases we present in this study are not currently implemented in other projects or publicly available.

\section{Methodology}

\begin{figure*}
	\centering
		\includegraphics[scale=0.6]{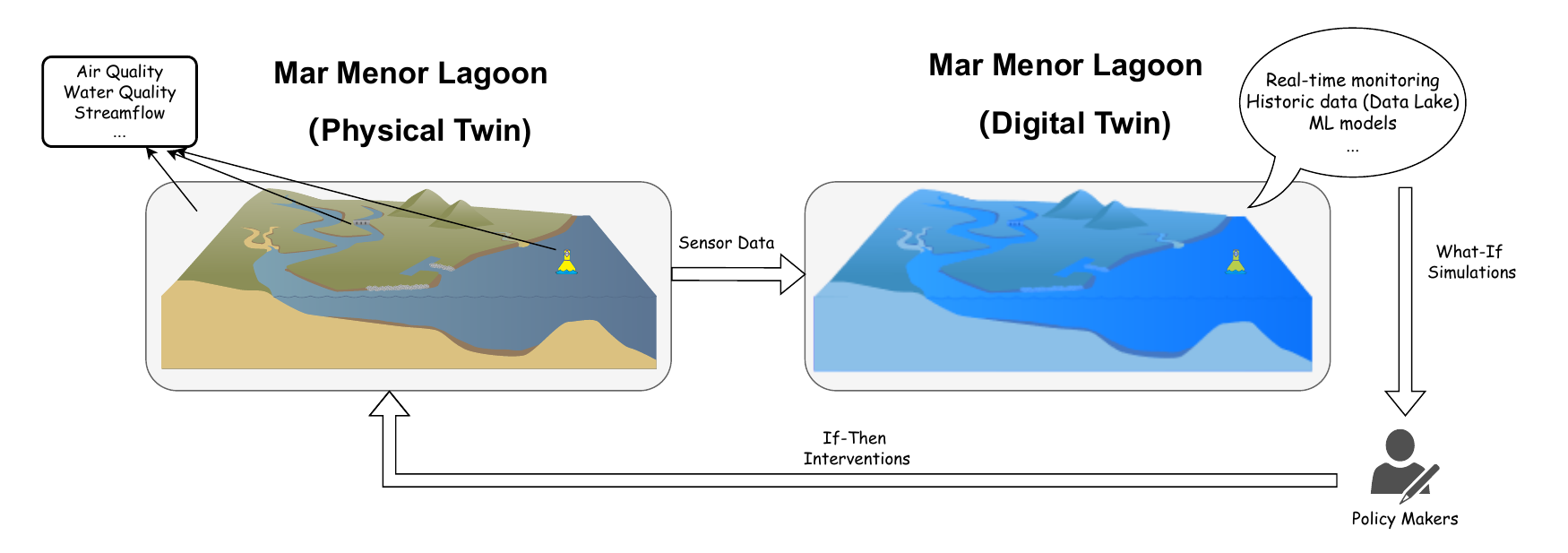}
	\caption{Overview of the DT background.}
	\label{FIG:DTBackground}
\end{figure*}

\autoref{FIG:DTBackground} shows an overview of the use of the DT. The DT developed consists of a web-based application and built with a three-tier architecture, namely data, application and client or presentation tier \citep{helu2017reference}. The data layer is the lowest layer in the architecture and its function includes collecting, organizing, and accessing data for use in other layers. The application layer is where the back-end of the DT is implemented. It provides an interface between the data layer and the presentation layer and implements the business logic of the application. Finally, the main function of the presentation layer, where the front-end is implemented, is to dynamically display the information stored in the data layer and interact with end users.

\subsection{System architecture overview}
\begin{figure*}
	\centering
		\includegraphics[scale=0.45]{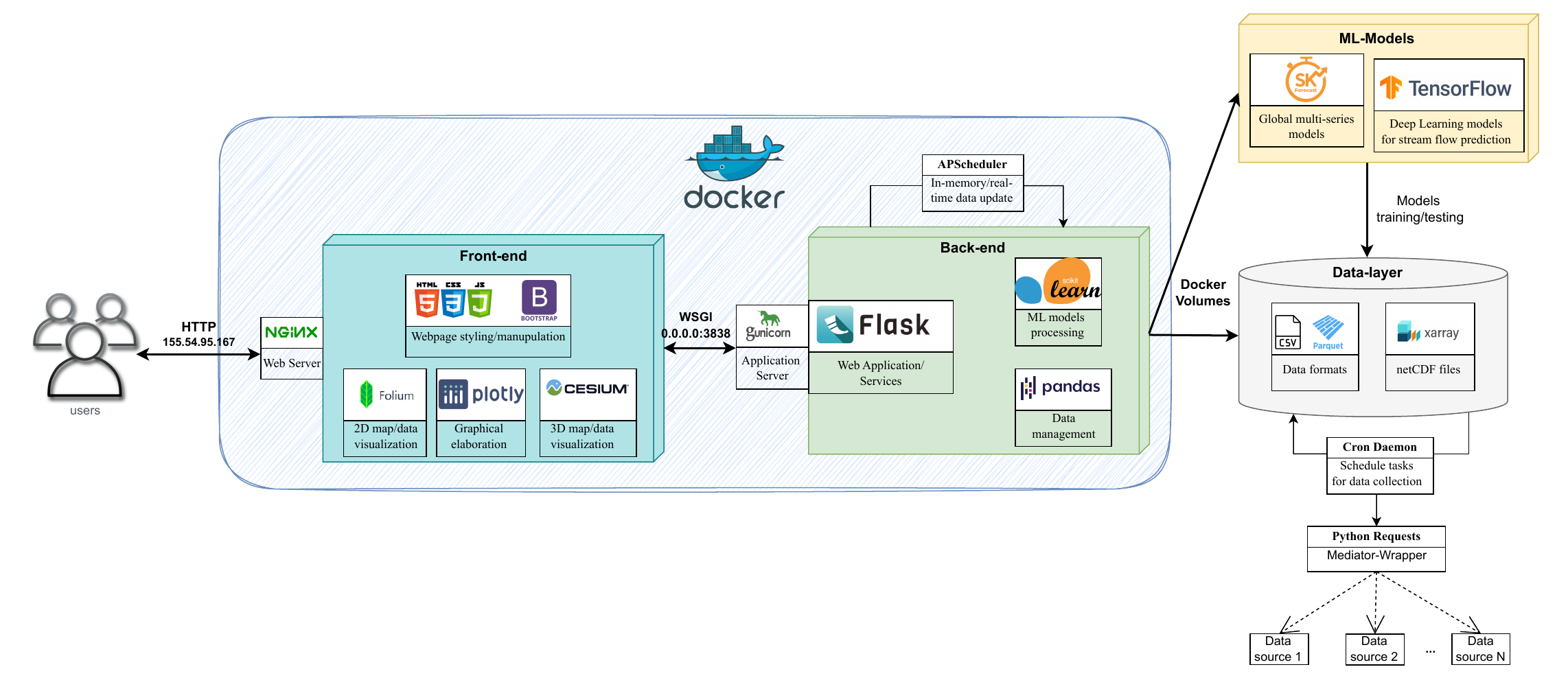}
	\caption{Architecture of the digital twin.}
	\label{FIG:archi}
\end{figure*}

\autoref{FIG:archi} shows the current architecture of the platform and the main components of each layer. It is important to note that all libraries and tools shown in the figure, and therefore used in our DTO, are open source.

\begin{figure*}
	\centering
		\includegraphics[scale=0.205]{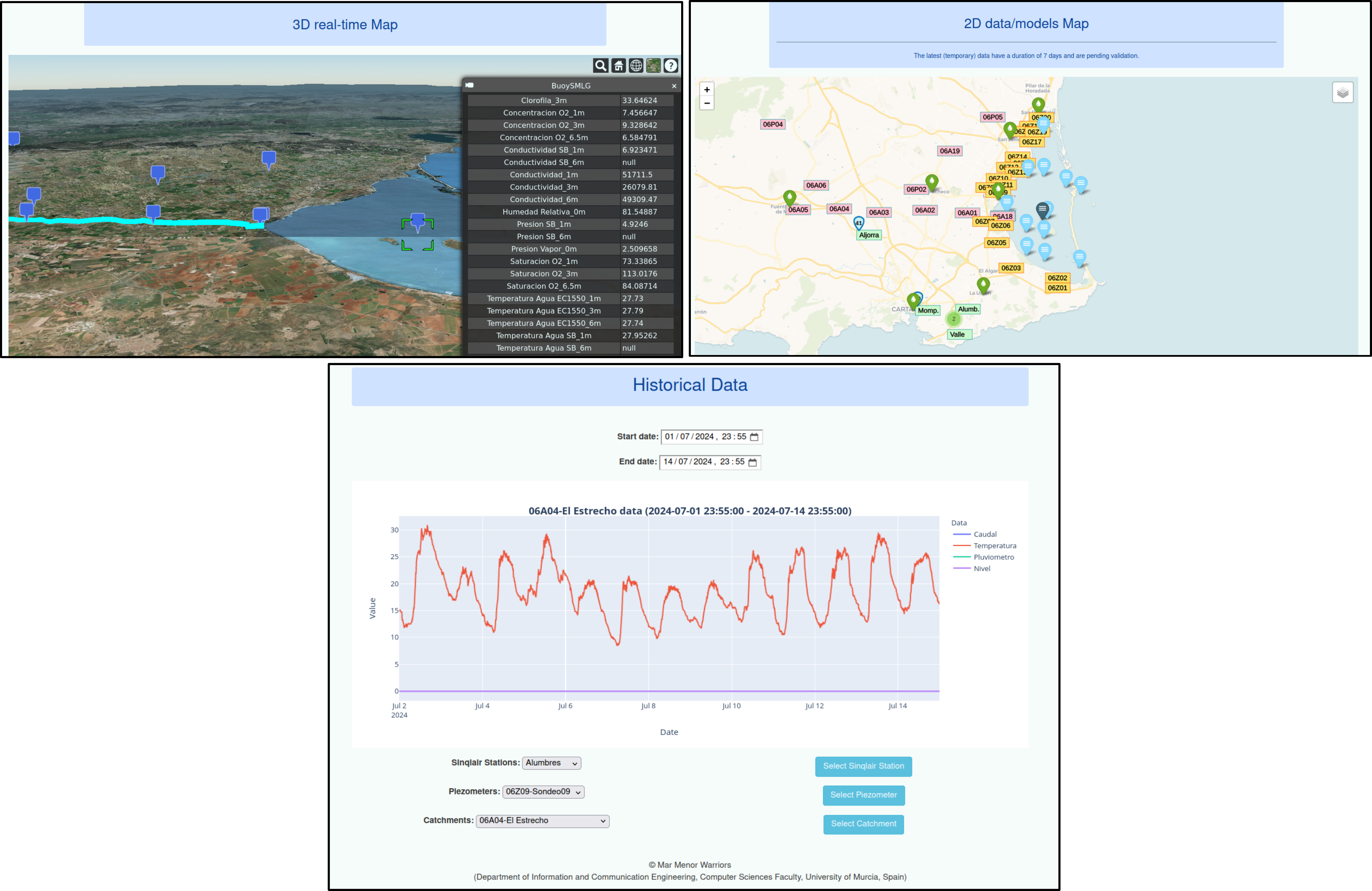}
	\caption{Overview of the developed DT app, which is divided into three sections. The first one shows the latest real-time data of each location in a 3D map. The second consists of a 2D map where users can visualize last week's data and the output of implemented models. The last one allows users to visualize long-term historical data.}
	\label{FIG:DT App}
\end{figure*}

In the front-end, the combination of HTML, CSS, and JavaScript is used for web page design, styling, and interactivity. The Bootstrap library\footnote{https://getbootstrap.com/} is also used to achieve a responsive web design, i.e. to adapt to different device sizes. Plotly\footnote{https://plotly.com/}, a browser-based graphing library has been used to plot interactive graphs. An important factor in any DT is the visualization of the physical twin. In our case, it corresponds to the entire Mar Menor Lagoon and its surrounding catchment basin. Therefore, both 2D and 3D maps are implemented as our virtual twin to represent the environment. For the 3D map, Cesium, an open platform for accurate 3D geospatial visualization, is chosen to display real-time data from various sources. On the other hand, Folium\footnote{https://python-visualization.github.io/folium} is the Python wrapper of Leaflet.js\footnote{https://leafletjs.com/}, one of the most widely used map visualization tools and has been chosen to plot an interactive 2D map that connects to Plotly to visualize the last seven days of data and also the output of ML models at each location. An overview of the developed DT Web App is shown in \autoref{FIG:DT App}.

In the back-end, the main core of this layer is Flask, a micro-web-framework with great capability and flexibility to create RESTful APIs using a simple routing/debugging system. Flask provides the interconnectivity for multiple microcomponents, ranging from scientific computing to data management to graphical interfaces \citep{bonney2022development}. In our DT, Flask is tasked with receiving the request sent from the front-end and processing it by calling Scikit-learn\footnote{https://scikit-learn.org/} (for ML modeling tasks, e.g., loading the model and making the prediction) and Pandas\footnote{https://pandas.pydata.org/} (for data management, e.g., loading the requested data and transforming it into a specific format). Both components have access to the ML-models and the data-layer to get the necessary resources for the task they are assigned. To keep the latest real-time data, APSCheduler\footnote{https://apscheduler.readthedocs.io} is used and it will reload (hourly) the data stored in memory. Another important feature of Flask is that it allows multithreading, meaning that the platform can handle multiple client requests concurrently, thus improving the performance and scalability of the DT.

The data layer, as mentioned earlier, is responsible for collecting and storing the heterogeneous data types. The Cron Daemon, a Unix operating system job scheduler, is used to periodically launch data collection tasks that follow the Mediator-Wrapper strategy (see \autoref{DC}). The developed models (see \autoref{Minimal models} and \autoref{DLmodels}) are stored in ML models and are ready for use. However, since the data is constantly updated, the models are periodically retrained to take advantage of the newly collected data and to improve their performance.

Finally, the DT is deployed on a VM running O.S. Ubuntu 22.04.2 LTS implemented on QEmu with 8 vCPU and 68 GB of RAM on a host with a Xeon Silver 4214R CPU @2.4GHz and is publicly accessible with the URL \url{http://155.54.95.167/}. The NGINX\footnote{https://nginx.org/} is selected as the HTTP and reverse proxy server to communicate with the client. It will serve static content directly from the front-end and will handle dynamic content by proxying the client's requests to the application server, which in this case is Gunicorn\footnote{https://gunicorn.org/}. Gunicorn is a Python Web Server Gateway Interface (WSGI) HTTP server designed to serve Python web applications. After receiving the request from NGINX, Gunicorn will translate it into a WSGI-compatible format and then call and execute the Python code (in Flask) and return the results when it is done. Both the front-end and back-end are packaged in Docker containers, and the Docker volumes provide access to the external data/models.

\subsection{Data collection}\label{DC}

In order to be able to simulate different \textit{what-if} scenarios, where each of them can require large amounts and diverse types of data from multiple domains, the DT contains the implementation of an initial version of a data lake and follows the Mediator-Wrapper or Mediator-based architecture \citep{duque2022building} to collect and store the heterogeneous types of data.

The data of the DT are mainly about the Mar Menor, located in the region of Murcia, Spain. All of them are publicly available. At the moment, it contains the following data from different sources and institutions: 
\begin{itemize}
    \item Catchments sensors and piezometers data from the Automatic Hydrological Information System (SAIH) of the Segura River Hydrographic Basin (https://\url{www.chsegura.es/es/cuenca/redes-de-control/saih/ivisor/}).
    \item Air Quality stations from Regional Ministry of Environment, Universities, Research and Mar Menor (\url{https://sinqlair.carm.es/calidadaire/}).
    \item Air Quality Index (AQI) from the World Air Quality Index (WAQI) project (\url{https://waqi.info/}).
    \item Buoy data from the SmartLagoon project (\url{https://marmenorsensing.com/applications/eb0sVoYBcSC7WUER9weF}).
    \item Sea environmental variables from Mar Menor Scientific Data Server of Technical University of Cartagena (UPCT) (\url{https://marmenor.upct.es/}).
    \item  Conductivity, Temperature and Depth (CTD) stations of Murcia Agricultural and Environmental Research and Development Institute (IMIDA) (\url{https://idearm.imida.es/cgi/siomctdmarmenor/}).
    \item Weather (real and prediction) data from the Spanish State Meteorological Agency (AEMET) (\url{https://www.aemet.es/es/datos_abiertos/AEMET_OpenData}). 
\end{itemize} 

\autoref{tab:dataSources} summarizes the source, parameters and granularity of data collected and to be used in DT.

\begin{table*}[]
\caption{Source, field, start date, parameters, temporal and spatial granularity of the collected and available data. Data is, to the date of publication of this manuscript, still being collected. 
}
\label{tab:dataSources}
\resizebox{\textwidth}{!}{%
\begin{tabular}{@{}lllllll@{}}
\toprule
Source & Field \& Area & Start date & Parameters & \begin{tabular}[c]{@{}l@{}}Data\\      Granularity\end{tabular} & \begin{tabular}[c]{@{}l@{}}Spatial \\      Coverage\\      (KM\textsuperscript{2})\end{tabular} & \begin{tabular}[c]{@{}l@{}}Observation\\      Points\end{tabular} \\ \midrule
SAIH (Catchments) & Coastal, River Basin & 2016-01-08 & Temperature (°C), Streamflow (m³/s), Water Level (m), Rain   Gauge (mm) & 5 minutes & 423.12 & 13 \\
SAIH (Pizometers) & Coastal, River Basin & 2019-12-13 & \multirow{2}{*}{\begin{tabular}[c]{@{}l@{}}Temperature (°C),   Conductivity (µS/cm), Piezometric Level (msnm), Salinity (PSU),\\      Total Dissolved Solids (mg/l)\end{tabular}} & 5 minutes & 41.06 & 19 \\
 &  &  &  &  &  &  \\
SINQLAIR & Air, River Basin & 2022-01-01 & O3, NO, NO2, NOx, NH3, NT, CO, SO2, PM10, Benzene, Toluene,   Xylene (all units in µg/m\textsuperscript{3}) & Hourly & 2544.02 & 8 \\
WAQI & Air, Spain & Only real time & Air Quality Index (AQI) of multiple air pollutants & Hourly & 1079347.19 & 382 \\
SmartLagoon & Marine, Lagoon & 2022-10-14 & \multirow{4}{*}{\begin{tabular}[c]{@{}l@{}}Air temperature (°C),   Relative Humidity (\%), Vapor Pressure (kPa), Wind Speed (m/s),  \\      O2 Concentration ($\mu$g/L), O2 Saturation (\%), Conductivity ($\mu$S/cm),   Chlorophyll ($\mu$g/L), \\      Water Temperature (°C), Water Temperature SB (°C), Water Temperature EC   1550 (°C),\\      Conductivity SB (S/m), Pressure SB (MPa), Turbidity (FTU), Depth (m)\end{tabular}} & 5 minutes & Single Point & 1 \\
 &  &  &  & or hourly &  &  \\
 &  &  &  &  &  &  \\
 &  &  &  &  &  &  \\
SDC-UPCT & Marine, Lagoon & 2017-03-07 & \multirow{3}{*}{\begin{tabular}[c]{@{}l@{}}Temperature (°C),   Salinity (PSU), Transparency (m), Chlorophyll (mg/m3 ), Oxygen (mg/l), \\      Turbidity (FTU), Phycoerythrin (ppm),    Chromophore Dissolved Organic Matter (ppb), \\      Depth (m)\end{tabular}} & $\approx$Weekly & 96.91 & 12 \\
 &  &  &  &  &  &  \\
 &  &  &  &  &  &  \\
CTD-IMIDA & Marine, Lagoon & 2016-06-08 & \multirow{2}{*}{\begin{tabular}[c]{@{}l@{}}Temperature (°C), Chlorophyll (mg/m3), Conductivity (S/m), Oxygen (mg/l), Salinity (PSU),\\      Transparency (m),  Turbidity (FTU), Organic materials (ppb), pH, depth (m)\end{tabular}} & $\approx$Weekly & 96.91 & 12 \\
 &  &  &  &  &  &  \\
AEMET & Weather, River Basin & 2023-07-26 & Air temperature (°C), Relative humidity (\%), Precipitation (mm) & Hourly & 327.35 & 7 \\ \bottomrule
\end{tabular}%
}
\end{table*}


Although all the data are collected from public institutions, since there is no standard method to retrieve information from each source, each dataset is different in its collection and format. Therefore, several Python scripts are implemented to download and organize the received data, following the Mediator/Wrapper architecture. 

To download the data, first, if the source entity has implemented an Application Programming Interface (API) to receive data, the process is simple. It is only necessary to set the parameters to request the needed data, such as the date range, the variables to download, and the localization information. However, if the request data is embedded in the source HTML file, web crawler techniques are used to extract the information of interest. This method involves a number of web technologies such as cookies, sessionID and regular expressions that adapt the structure of each web page. Finally, if the source offers only downloadable files, the solution is to retrieve the files (manually or via HTTP requests) and process them locally.

In our DT scenario, SINQLAIR, WAQI, SmartLagoon and CTD-IMIDA provide the corresponding API for data retrieval purposes, SAIH's records are collected using the web crawler method, and in SDC-UPCT, the datasets are in NetCDF format, retrieved via HTTP requests.

On the other hand, providing real-time or near real-time data is one of the most important functionalities of any DT. Most target sources fulfill this requirement by publishing their data in near real-time. Therefore, all implemented Python scripts are hosted in the server and executed according to each source's publishing schedule. 

One aspect that must be considered in all platforms, particularly those that provide data in real time, is the level of validation and error detection that said data has \citep{gonzalez2022intrinsic}. In that sense, there are also two types of scripts used to store either temporary or historical data. The first runs hourly or daily to collect the last published data in the temporary files, while the second is executed weekly to add the last week's data to the historical files. The main reason for this design is that if the data is published in real-time, as many of the target sources have stated, it is usually invalidated at the moment of publication, meaning that it may contain errors due to sensor malfunction or internal calculation. In addition, to speed up the read and write operations of temporary files, they are stored in CSV format and handled with the append mode of the Pandas library, which does not need to read the entire file to add new entries. However, in order to make better use of storage space, the historical files are saved in Parquet format, which has the ability of data compression and encoding schemes with improved performance to handle complex data types \citep{belov2021choosing}. Therefore, temporary files contain real-time data with a date range of the last seven days and with the possibility of inaccurate observations, while historical files include all validated data.

\subsection{Data interoperability}

In a complex DT, it is often faced with large volumes of heterogeneous data sources that have their own attributes, attribute types and relationships \citep{conde2021modeling}. The structure of the data that is used in a DT needs to be understood, interpreted, and utilized across disparate systems or platforms, regardless of their underlying technologies or architectures. In that sense, we wanted to provide our DT with data interoperability. Data interoperability enables different components of a system to communicate, share, and integrate data effectively, without encountering compatibility issues or data format inconsistencies.
To this end, we use the FIWARE open source platform, which aims to promote interoperability between different systems and services by providing a set of open standards, APIs and data models, and is a real reference option for the development of DTs in any domain \citep{conde2021modeling}.


More specifically, a data modeling implementation based on Next Generation Service Interfaces-Linked Data (NGSI-LD) using the SAIH and CTD-IMIDA datasets is performed as a proof of concept for future extension of the approach. NGSI-LD is specifically designed to improve data interoperability and enable seamless integration and interaction between digital twins and other systems by providing semantic interoperability, dynamic context management, and query subscription mechanisms amongst other functionality. 

\begin{figure*}
	\centering
		\includegraphics[scale=0.4]{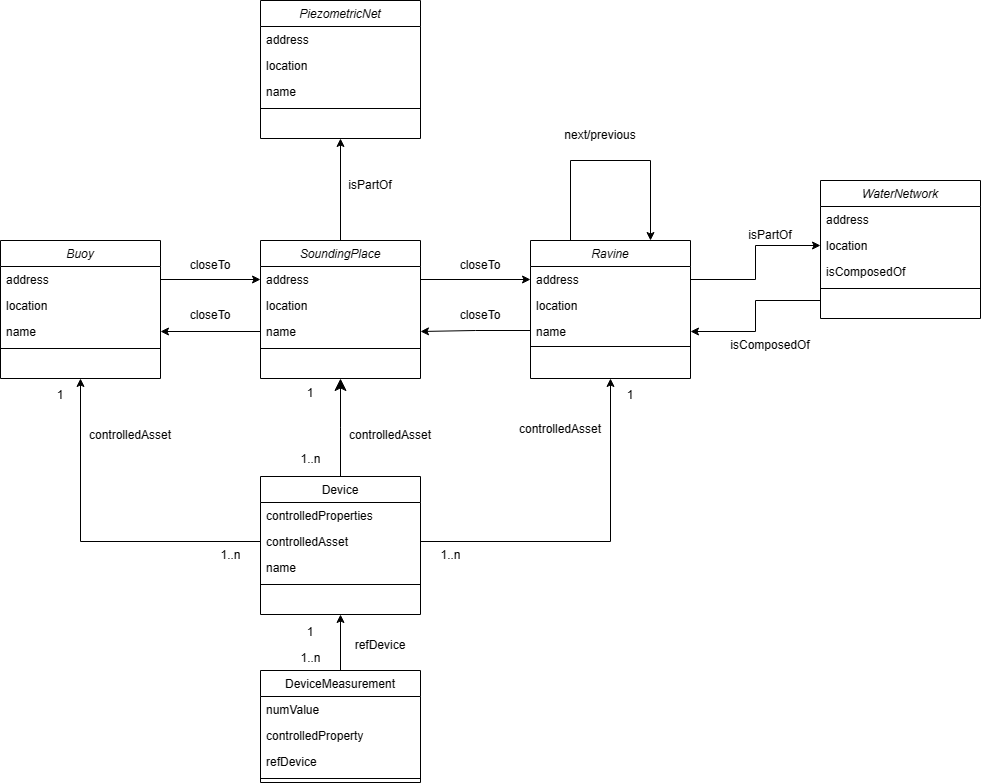}
	\caption{Data modeling of SAIH and CTD-IMIDA datasets.}
	\label{FIG:NGSI-LD-MODELS}
\end{figure*}

The data modeling schema used for the selected datasets (\autoref{FIG:NGSI-LD-MODELS}) is based on the FIWARE Smart Data Models project (\url{https://smartdatamodels.org/}), which has already standardized data models for many domains with the goal of being portable across different solutions. It is also possible to extend these models to fit the specific use case. The three main context entities are the buoy, PiezometricNet, and WaterNetWork, which represent the marine buoy, piezometers and ravine (or the stream gauge area) data, respectively. Both PiezometricNet and WaterNetWork consist of a set of locations where a piezometer or stream gauge area is located. For each location, its parameters are mapped by different devices, each of which contains historical data stored in DeviceMeasurement.

\begin{figure*}
	\centering
		\includegraphics[scale=0.63]{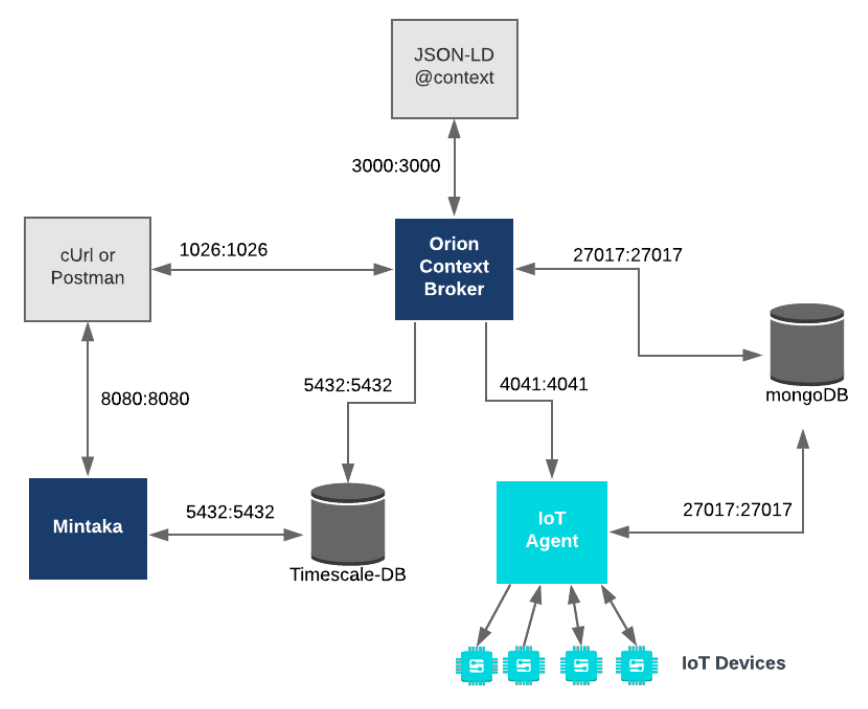}
	\caption{Architecture used to deploy the services (with docker). Image extracted from \url{https://github.com/FIWARE/tutorials.Short-Term-History/tree/NGSI-LD}}
	\label{FIG:NGSI-LD-ARC}
\end{figure*}

\autoref{FIG:NGSI-LD-ARC} shows the architecture used to provide the services (running in docker containers). The main components are: i) Orion Context Broker: It stores the current information from the various entities we have in NGSI-LD format; ii) MongoDB: Where the Orion Context Broker data is stored; iii) Mintaka: It is responsible for storing historical data. Automatically, when the Orion Context Broker entities are modified, it stores the modified data in its associated database along with a time stamp. iv) TimeScale-DB: Time series database where Mintaka will store the historical data of our entities.

After the raw data is converted/wrapped into NGSI-LD format, all this information is transformed into linked data that supports formal relationships, semantics, and property graphs (using the features offered by JSON-LD). The main advantage of this change is that all this information is now a machine-readable web of data, allowing access to new sources of information by forming knowledge graphs, for example, it is now possible to formalize queries based on geographic information of each entity through HTTP(s) requests: 
\begin{lstlisting}
curl -G -X GET 'http://localhost:1026/ngsi-ld/v1/entities'
-H 'Link: <https://raw.githubusercontent.com/yuye188/MarMenorDT/main/datamodels.context-ngsi.jsonld>; rel="http://www.w3.org/ns/json-ld#context"; type="application/ld+json"' 
-H 'Accept: application/json' 
-d 'type=Device' 
-d 'geometry=Point' 
-d 'coordinates=[37.7544,-0.8586]' 
-d 'georel=near;maxDistance==1000' 
-d 'options=keyValues'
\end{lstlisting}
\noindent this query returns the registered devices that are within 1 km of the coordinate [37.7544,-0.8586]:
\begin{lstlisting}
[{      
        "id": "urn:ngsi-ld:Device:015",
        "type": "Device",
        "alternateName": "Multiple sensors for Sounding Place 06Z11",
        "areaServed": "Mar Menor",
        "controlledAsset": [
            "urn:ngsi-ld:SoundingPlace:003"
        ],
        "controlledProperty": [
            "tds",
            "conductivity",
            "piezometricLevel",
            "salinity",
            "temperature"
        ],
        "dateLastValueReported": "2024-06-02T23:55:00Z",
        "description": "Device from Piezometric Net, belonging to the Sounding Place 06Z11",
        "deviceCategory": "sensor",
        "address": {
            "addressCountry": "ES",
            "addressRegion": "Murcia",
            "addressLocality": "Avenida de Munoz Zambudio, Los Alcazares, Cartagena",
            "postalCode": "30710"
        },
        "source": "https://saihweb.chsegura.es/apps/iVisor/visor_variable.php?punto=06Z11E10",
        "location": {
            "type": "Point",
            "coordinates": [
                37.7543,
                -0.8588
            ]
        },
        "name": "Device 015 found in 06Z11"
}]
\end{lstlisting}

Nevertheless, compared to the previous parquet-based methodology, the main drawback is the storage usage. For data of the same duration, the parquet-based method uses 328 MB of storage, while this NGSI-LD architecture uses 36.89 GB in Timescale DB and 331.2 MB in MongoDB (both as local volumes in Docker).

\section{Models}

\subsection{Minimal predictive models}\label{Minimal models}
In the context of a coastal ecosystem, identifying trends and accurately predicting variables across different domains in the short to medium term are crucial for providing valuable insights that inform and enhance environmental planning activities. To this end, an important functionality of our DT is to predict the next hours or day values for all variables available in each dataset.

However, as explained in previous sections, data from various domains is collected to develop different \textit{what-if} scenarios. Additionally, each domain typically includes data from multiple locations, with variables at each location potentially exhibiting distinct patterns. Consequently, applying the traditional Machine Learning (ML) train-test method to each variable is impractical in terms of efficiency and resource allocation.

To solve this problem, we have developed global models that provide N:M (multiple input, multiple output) features, i.e. modeling and predicting the future values of several time series at once. A multi-step forecast is performed using forecasters implemented in the Skforecast library \citep{skforecast}, which is specialized to handle ML processes applied to time series. The global models can reduce the potential noise that each series might introduce by capturing the core patterns that govern the series. While potentially sacrificing some individual insights, this approach is computationally efficient, easy to maintain, and can yield more robust generalizations across time series.

More specifically, for each location of the SAIH catchments and SINQLAIR air stations, a global model is developed to make auto-regressive predictions for all available variables. The model-building process begins with a preliminary study to exclude variables with more than 50\% missing values and use linear imputation to estimate the missing values. A function is created to set the weight of observations that have imputed values to zero during the training process so that these observations will not affect the performance of the model. The dataset is then divided into training (70\%), validation (10\%), and test (20\%) sets according to its length. To find the most appropriate model for each location, a set of algorithms (LGBMRegressor, XGBoost, Catboost, and HistGradientBoostingRegressor) are tested together with Bayesian search (implemented in Optuna \citep{optuna_2019}) to find the optimum hyperparameters. Finally, backtesting (a special type of cross-validation that evaluates the performance of a predictive model by applying it retrospectively to historical data) is used to select the model with the best performance, using Mean Absolute Error (MAE) and Coefficient of Variation of the Root Mean Square Error (CVRMSE) as metrics. The error unit of the former is the unit used for each variable, while the latter expresses the error in percent. 

The same process is carried out for the SDC UPCT, but not for all the buoys, since the variation between them is very small. Therefore, the data from the buoy located in the center of Mar Menor (number six) are used to predict the variables most relevant to the what-if scenarios proposed in \autoref{proposed what-if scenarios}, namely salinity, chlorophyll, turbidity, oxygen and transparency, all at 2 m depth.

\begin{figure*}
	\centering
		\includegraphics[scale=0.5]{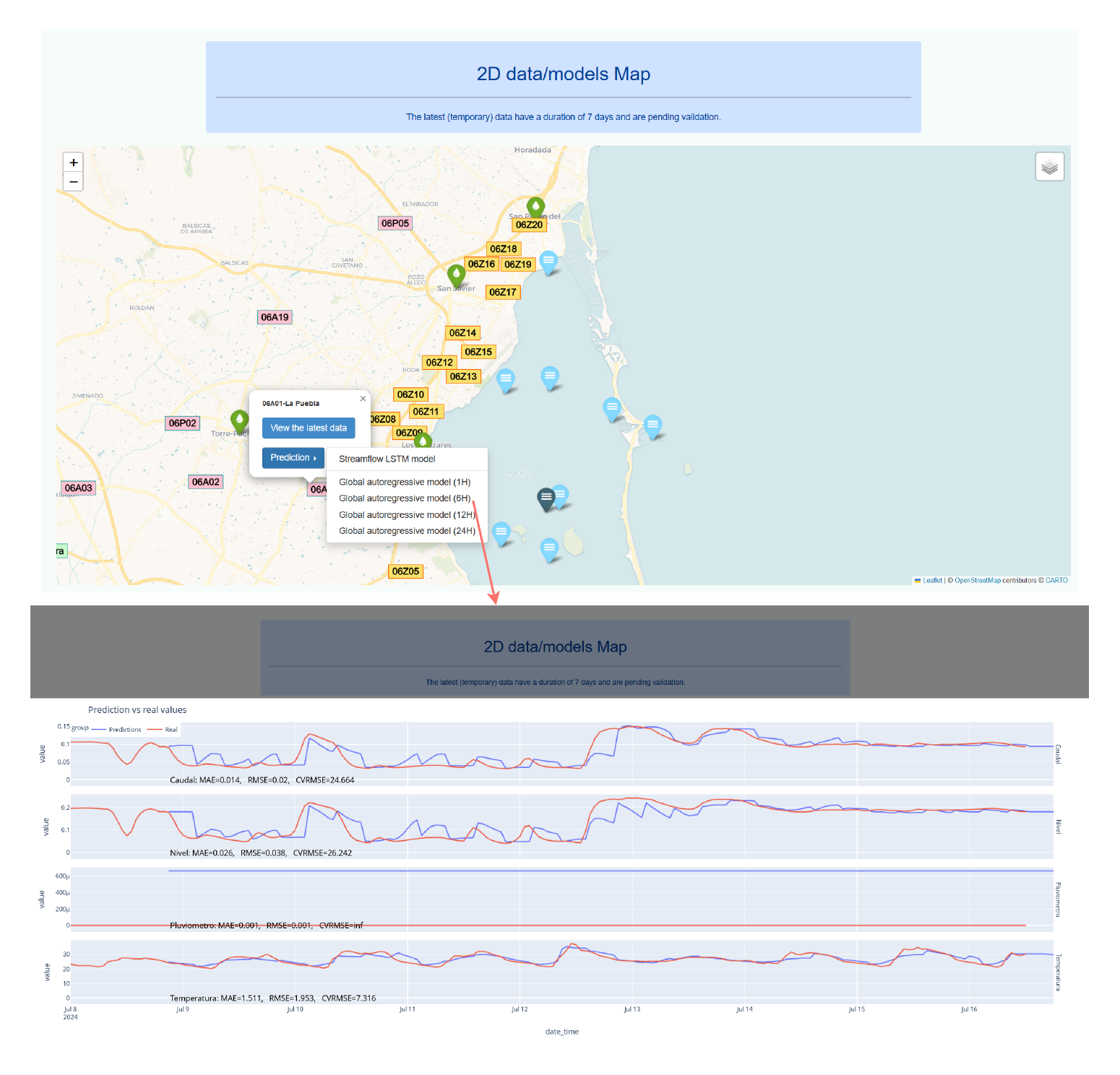}
	\caption{Showcase of the global auto-regressive models in a catchment stations (06A01-La Puebla) with a time horizon of 6 hours.}
	\label{FIG:showcaseGlobalModel}
\end{figure*}

Since there are a large number of variables, for simplicity, a subset of them, which can represent the advantages and shortcomings of using global models, is shown in \autoref{resultsGlobalModels}. As can be seen, CatBoost gave better results in the majority of cases of horizon between 1 and 24 hours and is therefore selected for use in the development of global models in these horizons. On the other hand, for horizons from 7 to 28 days, models using XGB had better performance in most of the cases, thus it will be used. It is also important to note that not all variables follow an auto-regressive pattern. This implies that if the output of a variable is not highly correlated with its previous values, e.g. streamflow or nitric oxide (NO), the global model will have a poor performance. 
Thus, these variables require further investigation, as in the case of the catchment streamflow variable presented in the next section. \autoref{FIG:showcaseGlobalModel} shows the results of the global model used for the catchment area (06A01-La Puebla) with a time horizon of 6 hours in our DT.

\begin{table*}[h]
\caption{Mean MAE and CVRMSE results of the CatBoost, HistGradientBoosting, LGBM and XGB models with 1, 6, 12, and 24 hours (for the parameters of the first block) and 7, 14, 21 and 28 days (for the parameters of the second block) forecast horizons. Only a subset of all available variables is shown.}
\label{resultsGlobalModels}
\resizebox{\textwidth}{!}
{\begin{threeparttable}
\begin{tabular}{@{}llcccccccc@{}}
\toprule
\multicolumn{1}{c}{Parameters} & \multicolumn{1}{c}{Horizon (h/d)} & \multicolumn{4}{l}{MAE} & \multicolumn{4}{l}{CVRMSE} \\ \midrule
 &  & \multicolumn{1}{l}{CatBoost} & \multicolumn{1}{l}{HistGradientBoosting} & \multicolumn{1}{l}{LGBM} & \multicolumn{1}{l}{XGB} & \multicolumn{1}{l}{CatBoost} & \multicolumn{1}{l}{HistGradientBoosting} & \multicolumn{1}{l}{LGBM} & \multicolumn{1}{l}{XGB} \\
\multirow{1}{*}{Temperature} & 1 & \textbf{0.617} & 0.653 & 0.650 & 0.704 & \textbf{4.782} & 5.046 & 5.095 & 6.279 \\
 & 6 & \textbf{1.343} & 1.545 & 1.429 & 1.615 & \textbf{10.489} & 12.215 & 12.259 & 16.265 \\
 & 12 & \textbf{1.789} & 2.239 & 1.966 & 2.218 & \textbf{13.627} & 16.661 & 15.013 & 20.600 \\
 & 24 & \textbf{1.904} & 2.165 & 2.038 & 2.416 & \textbf{14.256} & 16.345 & 15.439 & 25.983 \\
\multirow{1}{*}{Streamflow} & 1 & \textbf{0.003} & 0.003 & 0.003 & 0.004 & - & - & - & - \\
 & 6 & 0.833 & 0.669 & 0.070 & \textbf{0.010} & - & - & - & - \\
 & 12 & 0.893 & 0.530 & 0.617 & \textbf{0.016} & - & - & - & - \\
 & 24 & 0.861 & 0.359 & 0.650 & \textbf{0.022} & - & - & - & - \\
\multirow{1}{*}{Relative Humidity} & 1 & \textbf{2.797} & 2.798 & 2.878 & 2.988 & \textbf{6.414} & 6.448 & 6.606 & 6.793 \\
 & 6 & \textbf{5.902} & 6.164 & 6.203 & 6.278 & \textbf{13.238} & 13.767 & 14.640 & 14.019 \\
 & 12 & \textbf{7.979} & 9.083 & 9.473 & 9.235 & \textbf{17.284} & 19.276 & 32.466 & 19.866 \\
 & 24 & \textbf{9.036} & 9.607 & 9.942 & 10.060 & \textbf{19.338} & 20.119 & 31.843 & 23.711 \\
\multirow{1}{*}{Volume deposited} & 1 & \textbf{29.162} & 42.407 & 36.127 & 43.335 & \textbf{1.585} & 2.351 & 2.009 & 2.505 \\
 & 6 & \textbf{61.170} & 71.954 & 70.211 & 76.265 & \textbf{3.600} & 4.034 & 4.595 & 4.484 \\
 & 12 & \textbf{103.982} & 108.368 & 114.013 & 118.235 & 6.226 & \textbf{6.102} & 7.608 & 7.046 \\
 & 24 & \textbf{160.577} & 154.882 & 162.777 & 171.826 & 9.298 & \textbf{8.287} & 9.221 & 9.831 \\
\multirow{1}{*}{Wind Speed} & 1 & \textbf{0.607} & 0.615 & 0.616 & 0.619 & \textbf{30.795} & 31.065 & 31.109 & 31.126 \\
 & 6 & \textbf{1.006} & 1.060 & 1.064 & 1.069 & \textbf{49.126} & 51.820 & 51.949 & 52.255 \\
 & 12 & \textbf{1.147} & 1.225 & 1.228 & 1.257 & \textbf{56.238} & 60.594 & 60.879 & 61.628 \\
 & 24 & \textbf{1.293} & 1.375 & 1.379 & 1.421 & \textbf{64.090} & 68.724 & 69.059 & 69.069 \\
\multirow{1}{*}{CO} & 1 & \textbf{0.023} & 0.029 & 0.026 & 0.025 & \textbf{14.346} & 16.271 & 15.350 & 15.728 \\
 & 6 & \textbf{0.038} & 0.043 & 0.044 & 0.041 & \textbf{23.658} & 25.111 & 25.544 & 26.310 \\
 & 12 & \textbf{0.046} & 0.048 & 0.053 & 0.048 & \textbf{27.212} & 27.663 & 29.808 & 29.519 \\
 & 24 & \textbf{0.049} & 0.051 & 0.057 & 0.051 & \textbf{28.902} & 30.073 & 31.428 & 30.669 \\
\multirow{1}{*}{NO} & 1 & \textbf{1.980} & 2.059 & 2.105 & 2.151 & \textbf{101.866} & 105.218 & 104.879 & 106.968 \\
 & 6 & \textbf{3.233} & 3.442 & 3.631 & 3.744 & \textbf{142.321} & 153.736 & 148.170 & 148.397 \\
 & 12 & \textbf{3.660} & 3.928 & 4.640 & 4.403 & \textbf{149.548} & 153.593 & 161.707 & 155.164 \\
 & 24 & \textbf{4.117} & 4.271 & 5.967 & 5.002 & \textbf{158.522} & 160.417 & 181.673 & 161.649 \\
\multirow{1}{*}{O3} & 1 & \textbf{6.313} & 6.723 & 6.516 & 6.399 & \textbf{13.822} & 14.552 & 14.311 & 14.034 \\
 & 6 & \textbf{11.623} & 12.730 & 12.916 & 12.484 & \textbf{24.510} & 26.860 & 27.270 & 26.116 \\
 & 12 & \textbf{15.599} & 18.603 & 19.497 & 18.280 & \textbf{31.725} & 38.082 & 39.816 & 37.322 \\
 & 24 & \textbf{15.874} & 18.569 & 19.494 & 19.222 & \textbf{32.576} & 39.238 & 39.990 & 38.249 \\
\multirow{1}{*}{PM10} & 1 & 5.715 & \textbf{5.686} & 5.740 & 5.718 & 44.857 & \textbf{44.033} & 45.870 & 44.508 \\
 & 6 & 8.898 & \textbf{8.746} & 9.001 & 8.954 & 64.494 & \textbf{62.571} & 68.368 & 69.904 \\
 & 12 & 10.029 & \textbf{9.855} & 10.172 & 10.137 & 69.368 & \textbf{69.206} & 74.886 & 78.538 \\
 & 24 & \textbf{10.734} & 10.758 & 11.245 & 11.010 & 70.738 & \textbf{69.587} & 78.516 & 81.507 \\ \midrule
\multirow{1}{*}{Salinity} & 7 & \textbf{0.351} & 0.387 & 0.388 & 0.369 & \textbf{1.064} & 1.166 & 1.153 & 1.137 \\
 & 14 & \textbf{0.384} & 0.430 & 0.399 & 0.399 & \textbf{1.151} & 1.282 & 1.183 & 1.191 \\
 & 21 & 0.383 & 0.437 & 0.380 & \textbf{0.361} & 1.128 & 1.270 & 1.136 & \textbf{1.095} \\
 & 28 & 0.476 & 0.544 & 0.489 & \textbf{0.452} & 1.345 & 1.510 & 1.427 & \textbf{1.327} \\
\multirow{1}{*}{Chlorophyll} & 7 & 0.132 & \textbf{0.121} & 0.154 & 0.128 & 42.056 & \textbf{40.871} & 53.181 & 42.211 \\
 & 14 & 0.151 & 0.148 & 0.188 & \textbf{0.148} & \textbf{49.143} & 50.725 & 74.735 & 51.368 \\
 & 21 & 0.133 & 0.127 & 0.167 & \textbf{0.127} & \textbf{41.618} & 41.791 & 57.102 & 42.597 \\
 & 28 & \textbf{0.168} & 0.183 & 0.246 & 0.195 & \textbf{52.848} & 60.090 & 86.079 & 65.548 \\
\multirow{1}{*}{Turbidity} & 7 & 0.355 & 0.381 & 0.355 & \textbf{0.309} & 66.266 & 63.742 & \textbf{59.733} & 61.026 \\
 & 14 & 0.380 & 0.414 & 0.368 & \textbf{0.316} & 70.044 & 69.007 & 61.842 & \textbf{61.781} \\
 & 21 & 0.420 & 0.454 & 0.393 & \textbf{0.336} & 79.258 & 74.776 & \textbf{65.189} & 65.931 \\
 & 28 & 0.471 & 0.466 & 0.390 & \textbf{0.353} & 87.686 & 77.171 & \textbf{65.143} & 71.499 \\
\multirow{1}{*}{Oxygen} & 7 & 0.510 & 0.512 & 0.524 & \textbf{0.504} & 12.354 & 12.798 & 12.435 & \textbf{12.057} \\
 & 14 & 0.550 & 0.565 & 0.540 & \textbf{0.518} & 12.998 & 13.777 & \textbf{12.356} & 12.484 \\
 & 21 & 0.583 & 0.612 & 0.608 & \textbf{0.580} & 13.982 & 14.854 & 13.736 & \textbf{13.486} \\
 & 28 & 0.593 & 0.632 & 0.607 & \textbf{0.581} & 14.009 & 15.800 & 13.890 & \textbf{13.455} \\
\multirow{1}{*}{Transparency} & 7 & 1.381 & \textbf{1.098} & 1.154 & 1.108 & 23.746 & \textbf{19.228} & 20.241 & 19.554 \\
 & 14 & 1.544 & 1.305 & \textbf{1.138} & 1.259 & 26.556 & 22.797 & \textbf{20.306} & 22.221 \\
 & 21 & 1.707 & 1.448 & \textbf{1.210} & 1.335 & 29.620 & 25.467 & \textbf{22.326} & 23.911 \\
 & 28 & 1.723 & 1.467 & \textbf{1.173} & 1.292 & 30.084 & 26.022 & \textbf{21.482} & 23.021 \\ \bottomrule
\end{tabular}
\begin{tablenotes}
    \item  Note: Streamflow CVRMSE values are not shown because the range of streamflow values is very small, usually 0, so by poorly predicting the rises, their CVRMSE becomes huge (millions).
\end{tablenotes}
\end{threeparttable}}
\end{table*}

\subsection{Run-off what-if scenario: an integrated use case}\label{DLmodels}

In this section, we show the results of the what-if scenarios that have been implemented within the context of our DT. By analysing several hypothetical situations, we aim to discern the underlying dynamics within the Mar Menor ecosystem. Our approach is twofold: we demonstrate a subset of scenarios already included in the DT, while also introducing open what-if scenarios derived from contextual domain expertise that are not targeted in present but will serve for future investigation and reference to scale our DT.

\begin{figure*}
	\centering
		\includegraphics[scale=0.32]{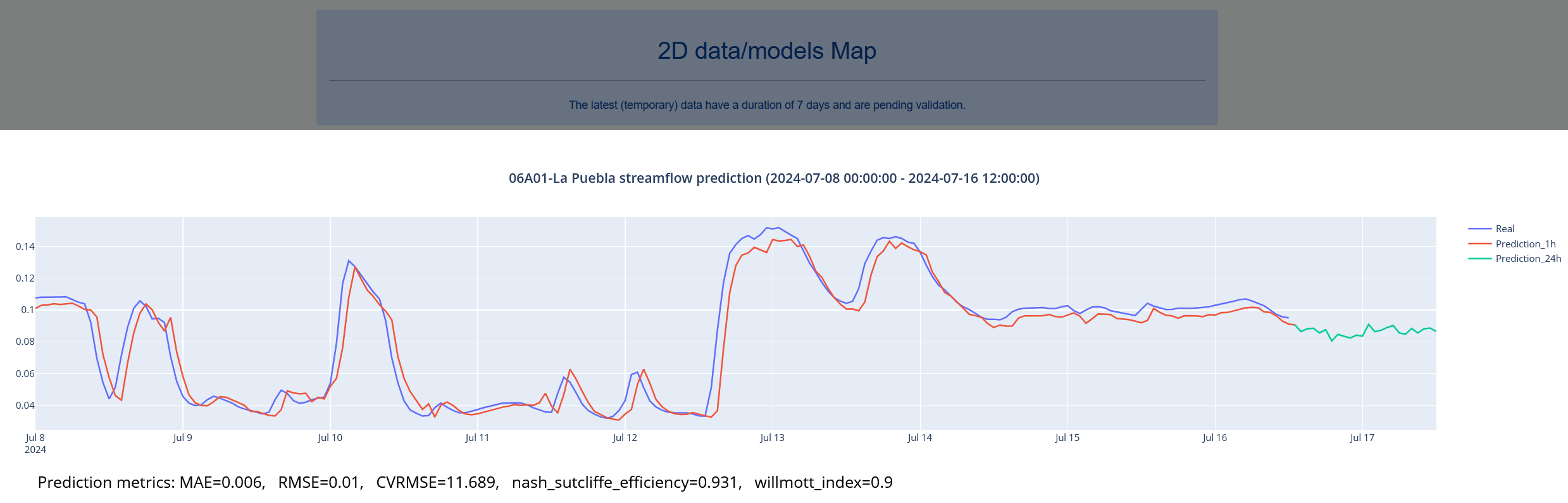}
	\caption{Showcase of the LSTM model in a catchment station (06A01-La Puebla) with different time horizons.}
	\label{FIG:showcaseLSTMModel}
\end{figure*}

For streamflow prediction in Mar Menor 
we developed an enhanced model using the Long Short-Term Memory network (LSTM) framework provided by Keras \citep{chollet2015keras}. Our optimization process involved experimenting with various LSTM configurations, ultimately identifying the most effective structure, which included: 1) an LSTM layer with 128 units and a tanh activation function; 2) a dense layer with 64 units and a linear activation function; 3) another dense layer with 32 units and a ReLU activation function; and 4) a final dense layer with a single unit and a linear activation function. We employed a RobustScaler for feature scaling to manage statistics that are robust to outliers effectively. Given the constraints of our dataset, where negative streamflow values are not practical (streamflow values are inherently non-negative), the model occasionally predicted negative values. To remedy this, we implemented a simple function to adjust negative predictions to zero. For practical application considerations, piezometer data, which cannot be collected hourly, was excluded from our model. However, we continued to utilize streamflow data from seven locations and rainfall measurements from ten rain gauge stations proximate to the watercourse. We developed predictive models for streamflow series at two points along the Albujón watercourse: La Puebla (06A01) and Desembocadura (06A18). The models utilized data collected from March 8, 2021, to April 20, 2024. For model training and evaluation, we partitioned the data into three subsets: 70\% was used for training, 10\% for validation, and the remaining 20\% for testing.  In the testing dataset for both locations, we achieved a CVRMSE of less than 11.50 for a 1-hour time horizon, 37.59 for a 6-hour time horizon, and 44.75 for a 24-hour time horizon. Ultimately, for operational deployment, we utilized the respective scalers and models saved from each experiment to generate hourly predictions for each time horizon. The output of the model implemented for 06A01-La Puebla is shown in \autoref{FIG:showcaseLSTMModel}.

In this what-if scenario, we hypothesized that utilizing precipitation forecast information from the Spanish State Meteorological Agency (AEMET) could enhance the accuracy of streamflow predictions in the watercourse. Precipitation is the primary factor influencing rainfall measurements at the rain gauge stations integrated into our models. Consequently, we incorporated precipitation, temperature, and humidity forecasts provided by AEMET from the seven nearest weather stations for the period from August 26, 2023, to June 30, 2024. We used the previous LSTM model including these AEMET forecasts to predict streamflow in the same two points. Therefore, with the inclusion of the AEMET forecast in this run-off what-if scenario we take advantage of the national state weather forecast to enhance the prediction. In the testing dataset for both locations, we achieved a CVRMSE of less than 8.05 for a 1-hour time horizon, 26.34 for a 6-hour time horizon, and 38.86 for a 24-hour time horizon.

\section{Mid-term prospects of the digital twin}

The large amount of goods and services provided by coastal lagoons (fisheries, pollutant retention, CO2 sequestration, bathing water quality, biodiversity) means that any imbalance has important socio-economic consequences and requires urgent management decisions \citep{newton2018assessing, velasco2018ecosystem, perez2020coastal}. One of the most important processes that directly or indirectly affect most of these goods and services is eutrophication, with its multiple consequences, including the proliferation of organisms, both as direct or secondary response to the inputs of nutrients or organic matter (macroalgae, phytoplankton (which can sometimes be toxic), jellyfish,  and the occurrence of hypoxia events with the consequent mortality of organisms. The triggers for this process are closely related to activities and processes in the watershed that lead to the entry of water of different salinity than that of the lagoon and excessive amounts of nutrients. The complex interaction of climatic, hydrographic, oceanographic and biological factors and the different spatio-temporal scales at which they operate make the monitoring of all the variables involved and the modeling of the processes involved difficult to comprehensively cover. It is therefore important that any digital twin, to be accurate, and sufficiently precise, considers at least the key variables and processes to address the various management needs and anticipate key problems.

How much freshwater can a coastal lagoon receive without significantly altering its functioning? What amount of nutrients trigger a dystrophic crisis? What is the difference between a continuous input and a massive discharge? When can torrential rain cause stratification of the water column? How long will it take for the system to recover after a DANA event? All those are questions that need further data collection and in the following subsections, we will review the potential additions to our Digital Twin in terms of data in order to solve different scenarios that are also proposed.

\subsection{Data integration}
In Subsection \ref{DC}, we have outlined the existing data sources integrated into the Digital Twin. Additionally, we have identified several data types that could enhance this information, enabling the creation of more simulation scenarios within the Mar Menor Digital Twin.

On the coastal side:
\begin{itemize}
\item Biochemical parameters of water in various streamflows: phosphates (mg PO\textsubscript{4}/l), nitrates (mg NO\textsubscript{3}/l), and conductivity ($\mu$S/cm).
\item Satellite images can be utilized to compute indexes that describe and monitor land use and health \citep{mendoza2024convolutional}, aiding in the evaluation of agriculture's impact on nearby streamflows.
\item Social data and community perspectives, including tourism rates. Including such information could facilitate informed decision-making and policy development that balances environmental conservation with economic development and tourism promotion.
\end{itemize}

Within the lagoon:
\begin{itemize}
\item Satellite images can generate maps indicating water salinity and temperature, chlorophyll concentration, turbidity, and other parameters using software like MAGO\footnote{\url{https://custom-scripts.sentinel-hub.com/sentinel-2/mago_water_quality_monitoring_tool/}}.
\item Manual sampling of physicochemical or biological variables such as pH, O$_2$, suspended solids, chl $\alpha$, silicates, ammonia, nitrates, nitrites, and phosphates allows for the assessment of water quality, nutrient availability, and primary productivity to calibrate remote sensing data or validate models.
\item Video data from boats and underwater robots can be processed to obtain detailed mappings and samplings of the seabed and algae presence, supporting underwater cartography, marine habitat assessment, and the detection of marine fauna such as fish and jellyfish, as well as environmental impact analysis.
\end{itemize}

Hydrological and hydrodynamic 3D models can provide essential insights into water dynamics for the watershed and lagoon relationship. Their outcomes can be valuable inputs for our AI-driven Digital Twin. 
These model outcomes can promote real-time decision-making, enabling proactive management of water resources, flood risk assessment, and optimization of water-related infrastructure and operations. At the same time, leveraging AI techniques within the DT enhances the accuracy and efficiency of hydrological modeling, allowing for adaptive strategies and improved resilience in the face of changing hydrological conditions. 

\subsection{Proposed what-if scenarios}\label{proposed what-if scenarios}

\subsubsection{Jellyfish blooms}

Will there be a jellyfish bloom in the next season, and how could changes in various factors affect it? For instance, how do seasonal thermal anomalies or prolonged heatwaves impact the life cycle of jellyfish species? Higher water temperatures in the lagoon could increase the likelihood of a bloom under certain conditions. Jellyfish proliferation in the Mar Menor often occurs in response to nutrient influx, exerting top-down control of the food web and maintaining water quality \citep{perez2002evidence, fernandez2020population}. However, predicting their dynamics remains challenging due to knowledge gaps in life cycle regulation \citep{fernandez2024unpredictability}. Jellyfish blooms are influenced by environmental factors such as water temperature, which triggers phase transitions in many species \citep{fernandez2020population, fernandez2023phenology}, salinity changes, and nutrient levels that promote algal growth—a key food source. Elevated chlorophyll-a and low dissolved oxygen levels can signal bloom conditions, while ocean currents and wind patterns influence jellyfish transport into lagoons. Predator-prey dynamics, food availability, and predation pressure also play roles. Integrating these factors into predictive models within a Digital Twin framework enhances bloom management and environmental monitoring.

As shown in Table \ref{resultsGlobalModels}, the predictions for several of the mentioned key variables are integrated into the DT for forecasting horizons of 1 to 24 hours and from 1 to 4 weeks. Temperature predictions achieve a CVRMSE ranging from 4.7 to 14.2 \% (1-24h), while wind speed predictions, which aid in understanding ocean currents, have a CVRMSE between 30\% and 60\% (1-24h), aligning with state-of-the-art wind estimation models \citep{yuan2018irregular, wang2021combined, groch2022forecasting}. In the case of salinity, we count with weekly data as shown in Table \ref{tab:dataSources}, and the accuracy is very high, having a CVRMSE between 1 and 0.4 \% (1-4 weeks). Nitrogen and phosphorus levels are not included due to the unavailability of public data, but they are closely linked to water inflow streamflow predictions discussed in Section \ref{DLmodels}. Similarly, we lack access to open data on jellyfish monitoring, which is necessary for developing accurate predictive models since they will be considered the output.

\subsubsection{Hypoxic crisis}
Coastal lagoons are highly sensitive to nutrient inputs, often leading to eutrophication \citep{thyssen1999nutrients, national2000clean}. This process has been documented in many lagoons, where it can become chronic and result in dystrophic crises with significant daily or seasonal fluctuations in primary production and oxygen levels \citep{reyes1991diel, boynton1996comparative, taylor1999responses}. During these crises, the ecosystem shifts between states of oxygen supersaturation, due to high autotrophic production and organic matter accumulation, and periods of anoxia characterized by high oxygen consumption \citep{d1997ecosystem, viaroli2001evolution, viaroli2004description}. Frequent anoxic conditions often lead to toxic phytoplankton blooms, mass die-offs of benthic organisms, and drastic changes in species distribution \citep{amanieu1975etude, bachelet2000seasonal, boutiere1982effet, ferrari1993planktonic, giusti2010assessment, reyes1991diel, sfriso1995nutrient, viaroli1996macrophyte, guyoneaud1998impact, hlailifate, specchiulli2010fluctuations}.
These events are more likely when factors increase oxygen consumption (such as organic matter influx, algal decomposition, and elevated temperatures) or reduce oxygen availability (such as temperature-driven decreases in oxygen solubility, water column stratification from freshwater influx, or periods of calm preventing mixing). Disruptions in the N/P ratio, particularly when phosphorus is no longer the limiting nutrient, are strongly associated with hypoxia in systems like the Mar Menor \citep{fernandez2022nutrient}. Freshwater and brackish water inputs further impact coastal lagoons by contributing suspended materials and nutrients, affecting eutrophication, water quality, salinity, hydrodynamics, species niches, and exchanges with the Mediterranean.

As shown in Table \ref{resultsGlobalModels}, we were able to successfully predict oxygen levels using the minimal models proposed in Section \ref{Minimal models}. In the different horizons, 1-4 weeks, the CVRMSE was always smaller than 0.6 \%. The estimations of oxygen levels will be later on correlated with other factors such as nutrient inputs and environmental conditions in order to better characterize and warn about possible anoxias and their impacts in the short and long term.

\subsubsection{Water quality (tourism and other activities)}
The Mar Menor is one of the few coastal lagoons capable of maintaining a high biological production, with a quality and transparency of its waters that make it ideal for bathing activities and water sports. Its average extinction coefficient k is below 0.45 (/m), with water transparency reaching 4.5 - 6 m Secchi disk visibility \citep{perez2019long}. However, various factors can affect water transparency, including eutrophication, coastal waves, geomorphology, rainfall, and coastal works. These turbidity events impact tourism appeal and have ecological consequences, affecting filter-feeding organisms and submerged vegetation. For instance, transparency losses have caused macrophyte meadow regression in deep areas and Caulerpa prolifera algae expansion in shallow areas \citep{garcia2012physiological, perez2012fisheries, perez2006changes}. Other factors like jellyfish protection nets can also deteriorate water quality. Predicting these events is crucial for management and regulation. While developing biological and hydrodynamic models can be complex \citep{chen2007nearshore, fischer2017spatio, madsen2001interaction, mao2012regional, singh2008simulation, stevenson2019advanced, stow2000deep, yamamoto2018analyzing}, machine learning and digital twin approaches integrating high-resolution data offer promising tools for anticipating and addressing these situations \citep{hafeez2019comparison, keiner1998neural, kim2023prediction, wang2021predicting}.

For the purpose of monitoring and predicting the water quality in the lagoon, we have included in our analysis the prediction of Turbidity and Transparency factors using the global models. In Table \ref{resultsGlobalModels} we can see that XGB provides the lower CVRMSE values, between 0.3 and 0.35 \% for Turbidity and in the case of Transparency the best results are obtained with LGBM algorithm giving a CVRMSE ranging between 1.1 and 1.2 for 1-4 weeks ahead estimation.

\subsubsection{Fishing Productivity}

Coastal lagoons are among the most productive ecosystems in the ocean and one of the main ecosystem services they offer are fishing yields and the quality of their products \citep{perez2012fisheries, newton2018assessing}. This productivity is determined by the intensity of the physical-chemical gradients they contain and the associated energy flows, in addition to the nutrient inputs that feed them \citep{perez2007hydrographic, perez2012fisheries, perez2024coastal}. However, fishing production can be affected by dystrophic crises and hypoxia events as well as by changes in the nature of the sediments, in macrophyte meadows or in factors that affect the recruitment or migrations of species from the open sea \citep{marcos2015long}. Fishing operations themselves and the marketing of their products can be affected by the proliferation of macroalgae and jellyfish eggs or toxic phytoplankton. The importance of the fishing sector in the regional economies around coastal lagoons and the uncertainties associated with the variability of all these parameters make it essential to have tools that allow us to understand the cause-effect relationships of the different variables and anticipate their consequences. 

\section{Conclusions and Future Work}

The development and deployment of Digital Twin (DT) technology for the Mar Menor coastal ecosystem is a significant advancement in environmental monitoring and predictive modeling. The comprehensive integration of diverse data sources, from biochemical parameters and satellite images to social data and hydrodynamic models, enables a holistic understanding of the ecosystem's dynamics. Our work illustrates the potential of DT to offer valuable insights for environmental management, particularly in predicting and mitigating issues such as eutrophication, hypoxia, jellyfish blooms, and fishing productivity.

Our efforts in creating predictive models for multi-step forecasting have demonstrated promising results. We have employed two primary approaches: global models and specific models. Global models predict variables that have similar characteristics, allowing for efficient and robust predictions across multiple time series. For these models, we utilize robust ML techniques, including Bayesian search for hyperparameter optimization and backtesting for model validation, achieving efficient and effective predictions. Conversely, certain variables require a more thorough approach due to their complexity (such as relating rain with streamflow). These specific models demand a more tailored and meticulous effort to obtain accurate results, addressing the distinct patterns and dependencies inherent to each variable.
Furthermore, the implemented FIWARE-based architecture has successfully demonstrated our DT's ability to model and provide data interoperability with large volumes of heterogeneous data. However, the high storage consumption of the application is the main drawback that should be addressed in future improvements.

As future work, in terms of the Digital Twin architecture:
\begin{itemize}
    \item Improve users' interaction with the DTO, allowing them to download and upload data and models.
    \item Implement an identification system with different roles.
    \item Include the framework Eclipse Ditto for the integration of physical systems, providing a scalable and flexible solution for modeling and simulation.
\end{itemize}
In terms of the Digital Twin models:
\begin{itemize}
    \item Integration with Hydrological and Hydrodynamic Models: Combining hydrological and hydrodynamic model outputs with AI-driven DT frameworks will provide a comprehensive understanding of water resource behaviors and responses to environmental changes.
    \item Development of specific models for variables that do not conform well to global modeling techniques, developing and refining specific models will be crucial. This will involve the creation of tailored solutions to accurately predict these variables, thereby improving the overall predictive capabilities of the DT.
    \item To incorporate advanced missing value estimation techniques such as the Bayesian-based ones \citep{gonzalez2020missing} and to include multivariate time series segmentation methods for general pattern extraction \citep{gonzalez2024multibeats}.
    \item To include warnings based on predictions in order to facilitate decision-making.
\end{itemize}
And finally, as a general goal, to engage with stakeholders, including local communities, policymakers, and environmental organizations since it is crucial for the success of our DT. By incorporating their perspectives and needs, we can ensure that the DT provides actionable insights that balance environmental conservation with socio-economic considerations. 

\printcredits

\section*{Declaration of Competing Interest}
\noindent The authors declare that they have no known competing financial interests or personal relationships that could have appeared to influence the work reported in this paper.

\section*{Acknowledgements}
This work forms part of the ThinkInAzul programme and was supported by MCIN with funding from European Union NextGenerationEU (PRTR-C17.I1) and by Comunidad Autónoma de la Región de Murcia - Fundación Séneca and 21591/FPI/21. Fundación Séneca, Región de Murcia (Spain). It was partially funded by the HORIZON-MSCA-2021- SE-01-01 project Cloudstars (g.a. 101086248) as well as by projects GEMINI TED2021-129767B-I00 and PERSEO PDC2021-121561-I00 both financed by MCIN/AEI /10.13039/501100011033 and the European UnionNextGenerationEU/PRTR.

\bibliographystyle{cas-model2-names}

\end{document}